\documentclass[preprint,showpacs,showkeys,preprintnumbers,amsmath,amssymb]
{revtex4}

\usepackage{amsmath}
\usepackage{amssymb}

\newtheorem{theorem}{Theorem}

\newtheorem{lemma}[theorem]{Lemma}

\newtheorem{proposition}[theorem]{Proposition}

\begin{document}

\title{Quantization and Asymptotic Behaviour of $\varepsilon
_{V^{k}}$ \\ Quantum Random Walk on Integers\footnote{Submitted to
Proc. of 3rd  NEXT Sigma-Phi Conference, Kolymbari 13-18 Aug.
2005, Eds. G. Kaniadakis, A. Carbone, M. Lissia}}
%
%
\author{Demosthenes Ellinas $^\dag$ and Ioannis Smyrnakis $^\ddag$ }
\address{
Technical University of Crete, \\
Department of Sciences, Division of Mathematics, \\
GR 731 00, Chania, Greece.\\
\ \ e-mail: $^\dag$ ellinas@science.tuc.gr \ \ $^\ddag$
smyrnaki@tem.uoc.gr }
\begin{abstract}
Quantization and asymptotic behaviour of a variant of discrete
random walk on integers are investigated. This variant, the
$\varepsilon _{V^{k}}$ walk, has the novel feature that it uses
many identical quantum coins keeping at the same time
characteristic quantum features like the quadratically faster than
the classical spreading rate, and unexpected distribution cutoffs.
\ A weak limit of the position probability distribution (pd) is
obtained, and universal properties of this arch sine asymptotic
distribution function are examined. \ Questions of driving the
walk are investigated by means of a quantum optical interaction
model that reveals robustness of quantum features of walker's
asymptotic pd, against stimulated and spontaneous quantum noise on
the coin system.
\end{abstract}
 \maketitle

\section{The $\protect\varepsilon _{V^{k}}$ Quantum Random Walk on Z}

Recent research activity on the topic of the so called quantum
random walks  has achieved a number of interesting results and
extensions of the usual notion of random walk (see e.g the
following articles and reviews
\cite{kempe,aharonov,meyer,kempe1,childs1,fahri2,ambainis3,ambainis4,konno2,
travaglione,sanders,dur,ellinas}). Our present study investigates
further quantum walks and their asymptotic statistical behaviour.
Let us consider a walker system with Hilbert space
$H_{w}=span\{|n>|\,n\in Z\}$, and a coin system with space
$H_{c}=span\{\left\vert +\right\rangle ,\left\vert -\right\rangle
\}$. \ Let the evolution operator be of the form
\begin{equation}
V=(P_{+}\otimes E_{+}+P_{-}\otimes E_{-})U\otimes \mathbf{1}.
\end{equation}%
This acts on the tensor product space $H_{c}\otimes H_{w}$. \ Here
$U$ is a unitary operator acting on the coin Hilbert space and
$P_{+},P_{-}$ are the projection operators in the distinguished
basis $\{\left\vert \pm \right\rangle \}$ of $H_{c}$. \ Also
$E_{\pm }$ are the right/left one step operators in the
distinguished basis $\{\left\vert n\right\rangle \;|\;n\in Z\}$ of
$H_{w}$. \ As they are commuting they share the eigenbasis
$\left\{ |\phi >=\frac{1}{2\pi }\sum\limits_{n\in Z}e^{-in\phi
}|n>|\phi \in \lbrack
0,2\pi ]\right\} ,$ of orthogonal elements viz. $<\phi |\phi ^{\prime }>=%
\frac{1}{2\pi }\delta (\phi -\phi ^{\prime }),$ which satisfy the eigenvalue
equations $E_{\pm }|\phi >=e^{\pm i\phi }|\phi >.$

The evolution operator $V$ acts on the state $|\phi \rangle$ to
give $V$ $|\phi\rangle=V(\phi )$ $|\phi \rangle$, where $V(\phi
)=\left( e^{i\phi }P_{+}+e^{-i\phi }P_{-}\right) U=e^{i\phi \sigma
_{3}}U.$\ Here $V(\phi )$ is a unitary operator that acts on
$H_{c}$. \ Suppose now that we consider the $\varepsilon _{V^{k}}$
model \cite{ellinas1, ellinas2}, in which the one-step evolution
of the walker density matrix $\rho ^{0}$ is given by $\rho
^{1}=\varepsilon _{V^{k}}(\rho ^{0})\equiv Tr_{H_{c}}(V^{k}\rho
_{c}\otimes \rho ^{0}V^{\dag k})$, where $ \rho _{C}$ is the coin
density matrix. \ Here $\varepsilon _{V^{k}}$ is a completely
positive trace preserving (CPTP) map that acts on $H_{w}\otimes
H_{w}^{\ast }$, however, since it has the property that it maps
density matrices to density matrices we can reduce its action to
the convex subset of density matrices $D(H_{w})\subset
H_{w}\otimes H_{w}^{\ast }$. \ Suppose that $\rho
^{0}=\int\limits_{0}^{2\pi }\int\limits_{0}^{2\pi }\rho (\phi
,\phi ^{\prime })|\phi ><\phi ^{\prime }|$. \ We have that
$\varepsilon _{V^{k}}(|\phi ><\phi ^{\prime }|)=Tr_{H_{c}}\left(
V^{\dag k}(\phi ^{\prime })V^{k}(\phi )\rho _{c}\right) |\phi
><\phi ^{\prime }|\equiv A(\phi ,\phi ^{\prime })|\phi ><\phi
^{\prime }|$, hence $\rho ^{1}=\int\limits_{0}^{2\pi
}\int\limits_{0}^{2\pi }\rho (\phi ,\phi ^{\prime })A(\phi ,\phi
^{\prime })|\phi ><\phi ^{\prime }|$. \ The $n-$step evolved
walker density matrix is \ $\rho ^{n}=\int\limits_{0}^{2\pi
}\int\limits_{0}^{2\pi }\rho (\phi ,\phi ^{\prime })A(\phi ,\phi
^{\prime })^{n}|\phi ><\phi ^{\prime }|,$ where $A(\phi ,\phi
^{\prime })$ is the \textit{characteristic function} of the
quantum random walk.

The \textit{position observable} $\ L|m>=m|m>,$ $m\in Z,$ and its
positive powers lend themselves to study the statistical moments
of the quantum walker after $n$ steps. To emphasize the connection
with the classical walk we also introduce a sequence of classical
random variables $L_{c}^{(n)}$ over the common sample space $Z,$
that correspond to the positions of the classical walker after $n$
steps, with probability distribution $
P(L_{c}^{(n)}=m)=Tr(|m\rangle \langle m|\rho ^{n})=\langle m|\rho
^{n}|m\rangle .$ Then we obtain for the statistical moments
\begin{equation}
\langle L^{s}\rangle _{n}\equiv Tr(L^{s}\rho ^{n})=\frac{1}{2\pi i^{s}}%
\int\limits_{0}^{2\pi }d\phi \left[ \partial _{\phi }^{s}\left[
\rho (\phi ,\phi ^{\prime })A^{n}(\phi ,\phi ^{\prime })\right]
\right] _{\phi ^{\prime }=\phi }= \sum_{m\varepsilon
Z}m^{s}P(L^{(n)}=m)\equiv \langle L_{c}^{s}\rangle _{n}
\end{equation}
A further study of the asymptotic behavior of the first and second moment
leads to the following

\begin{lemma}
The mean position of the quantum random walk is of the form $\mu
_{n}=\langle L\rangle _{n}=K_{1}n+\mu _{0}$ , while the variance is of the
form $\sigma _{n}^{2}=\langle L^{2}\rangle _{n}-\langle L\rangle
_{n}^{2}=K_{2}n^{2}+K_{3}n+\sigma _{0}^{2}$, where $K_{1},K_{2},K_{3}$
depend on the initial coin density matrix, the initial walker density matrix
and the tracing scheme, however they are independent of the number of
evolution steps taken, and \ $\mu _{0}$ and $\sigma _{0}^{2}$ are the mean
and variance of the initial position distribution. \ In particular, if the
quantum walker is initially in the at the position 0, that is $\rho
_{0}=|0><0|$, then $\mu _{n}=K_{1}n$ and $\sigma _{n}^{2}=K_{2}n^{2}+K_{3}n.$
\end{lemma}

Remark on $U-$quantization: Let us consider the general QRW as
introduced above, in the particular case where $U=\mathbf{1},$ we
have that $\ \varepsilon _{V^{k}}(\rho ^{0})=\rho
_{c++}E_{+}^{k}\rho ^{0}E_{+}^{\dag k}+\rho _{c-\text{
}-}E_{-}^{k}\rho ^{0}E_{-}^{\dag k}.$ This admits the
interpretation that the walker shifts its position $k$ steps to
the right (left) with probability $\rho _{c++}$ $(\rho _{c-\text{
}-})$. \ So if we are given a classical random walk on $Z$ \ with
nearest neighbor transition probabilities $p,1-p,$ we can select a
coin density matrix whose diagonal elements are these transition
probabilities in the preferred coin basis, and view in this way
the classical random walk as an $\varepsilon _{V^{k}}$ quantum
random walk with $U=\mathbf{1}$, and step size $k$. \ A natural
way then to quantize this classical walk is a unitary choice so
that $U\neq \mathbf{1}$. \ This amounts to a continuous
deformation of classical walk that may result into a new walk with
novel quantum features, as its known.\ Also since $U$ acts only on
coin space, and we also need to insert a consistent coin density
matrix, this $U-$\textit{quantization} procedure as may be called,
should be understood as a quantization of the coin of a classical
random walk. Generally the so obtained $U-$ quantized random walks
may admit solutions that employ off diagonal walker density
matrices, and these are considered to be genuine quantum
mechanical random walks. In fact the $U-$ quantization is not
restricted to walks on integers, but it can be used more generally
to quantized classical walks on other fields of numbers and
general lattice systems.

\section{Asymptotic Behaviour of Quantum Random Walk}

Next we study the behaviour of the walker when the number of steps $n$, is
large. \ In this case we have that
\begin{equation}
\langle L^{s}\rangle _{n}=\frac{n^{s}}{2\pi
i^{s}}\int\limits_{0}^{2\pi
}d\phi \rho (\phi ,\phi )\left[ \partial _{\phi }A(\phi ,\phi ^{\prime })%
\right] _{\phi ^{\prime }=\phi }^{s}+O(n^{s-1})\equiv \frac{n^{s}}{2\pi }%
\int\limits_{0}^{2\pi }d\phi \rho (\phi ,\phi )h(\phi
)^{s}+O(n^{s-1}). \label{asmom}
\end{equation}%
Here we have defined the \textit{asymptotic characteristic function} $h(\phi
)$ of the walk as%
\begin{equation}
h(\phi )=-i\left[ \partial _{\phi }A(\phi ,\phi ^{\prime })\right]
_{\phi ^{\prime }=\phi }=Im Tr_{H_{c}}(V^{k\dag }(\phi )\left(
V^{k}(\phi )\right) ^{\prime }\rho _{c}).  \label{HHH}
\end{equation}%
\ For the $\varepsilon _{V^{k}}$ model it reads
\begin{equation}
h(\phi )=Tr_{H_{c}}\left[ (\sigma +V^{\dagger }(\phi )\sigma V(\phi )+\cdots
+V^{\dagger k-1}(\phi )\sigma V^{k-1}(\phi ))\rho _{c}\right]  \label{HH}
\end{equation}%
where $\sigma =U^{\dagger }\sigma _{3}U$ is a rotated $\sigma _{3}$ Pauli
matrix. \

As the following theorem states the sequence of probability measures $\ \{%
\frac{1}{n}Tr(|m\rangle \langle m|\rho ^{n})=P(\frac{L_{c}^{(n)}}{n}=\frac{m%
}{n}),$ $n=1,2,...\}$ has all its moments converging to the moments of $%
Y=h(\phi ),$ as $n$ goes to infinity, where $\phi $ is stands for a random
variable (rv) on the circle with measure $\frac{1}{2\pi }\rho (\phi ,\phi ).$

\begin{theorem}
The sequence of classical random variables
$\frac{L_{c}^{(n)}}{n},n=1,2,...$ , corresponding to the sequence
of quantum observables, converges weakly to the random variable
$Y=h(\phi ),$ where $\phi $ is now a random variable with values
on the circle and measure $\frac{1}{2\pi }\rho (\phi ,\phi ).$ The
probability distribution function for $Y$ is given by the formula
\begin{equation}
P(y_{1}\leq Y\leq y_{2})=\frac{1}{2\pi }\int\limits_{y_{1}\leq
h(\phi )\leq y_{2}}\rho (\phi ,\phi )d\phi =\frac{1}{2\pi
}\sum_{i}\int \limits_{y_{1}}^{y_{2}}\rho
(h_{i}^{-1}(y),h_{i}^{-1}(y))\frac{1}{|h^{\prime
}(h_{i}^{-1}(y))|}dy,
\end{equation}%
where it is assumed that locally in the interval $[y_{1},y_{2}]$
the function $h$ admits a number of local inverses, labelled by
$i$. \
\end{theorem}

Remarks: \textit{i}) The dual $\varepsilon ^{\ast }(X)$ of a CPTP $%
\varepsilon (\rho )=\sum_{i}A_{i}\rho A_{i}^{\dagger }$ operating on a
observable $X$ and a density operator $\rho $ respectively is obtained to be
$\varepsilon ^{\ast }(X)=$ $\sum_{i}A_{i}^{\dagger }XA_{i}$ \textit{ii}) The
asymptotic characteristic function has a finite Fourier expansion, hence it
is locally invertible except from a set of measure zero, if the function is
not a constant. \ Further the inverse is differentiable except from a set of
measure zero. \ Hence the asymptotic distribution of $Y$ \ given above holds
as long as $h$ is not a constant. \ A somewhat similar theorem is obtained
in \cite{grimmett1} for the original QRW with two essential differences. \
First, in \cite{grimmett1} the sample space of rv $\phi $ was enlarge by
taking a product of the circle with a set of two points. \ Second, the
initial walker-coin state information was encoded in the measure while in
the present case only the initial walker state information is encoded in the
corresponding measure. \

It is now reasonable to ask what happens if $h$ is constant function. \ This
is answered by the following lemma:

\begin{lemma}
If the asymptotic characteristic function $h$ is constant then $\sigma
_{n}^{2}=\langle L^{2}\rangle _{n}-\langle L\rangle _{n}^{2}=K_{3}n+\sigma
_{0}^{2},$ which means the quantum random walk spreads classically. \
\end{lemma}

This can be proven simply by considering the asymptotic expansion
of the \ moments. As an example, let $|+>$ be the initial coin
state, and choose $ U=e^{i\frac{\pi }{4}\sigma _{2}}.$ \ The
walker state is initially $|0>,$ while the model is taken to be
the $\varepsilon _{V^{2}}.$ In this case it turns out that $h(\phi
)=-\cos (2\phi ),$ while $\rho (\phi ,\phi )=1.$ \

Remarks: \textit{i}) The range of values of $h(\phi )$ is the
interval $ [-1,1].$ Given that the possible values of walker
position after $n$ steps extends on the interval $[-2n,2n],$ the
normalized position asymptotic range of values is expected to be
the interval $[-2,2].$ This means that the asymptotic position
distribution experiences a sudden cutoff at half the spread one
would expect. \ This cutoff has also been observed with different
quantum evolution schemes \cite{ambainis4}, \cite{konno1},
\cite{konno2}, \cite{grimmett1}. \ \textit{ii}) Since $\phi $
takes values in \ the interval $[0,2\pi ],$ there are four
relevant inverses of $h$ with domain the interval $[-1,1],$ and
range in $[0,2\pi ].$ All these inverses satisfy the relation
$h^{\prime }(h_{i}^{-1}(y))=\pm 2\sqrt{1-y^{2}}.$ The resulting
asymptotic distribution is
\begin{equation}
P(y_{1}\leq Y\leq y_{2})=\frac{1}{\pi }\int\limits_{y_{1}}^{y_{2}}\frac{1}{%
\sqrt{1-y^{2}}}dy.  \label{asymp_pdf}
\end{equation}%
It is worth mentioning that this is the same distribution as the one
obtained in \cite{konno3}, in a different context. \

In fact a more general result about the asymptotic distribution
can be proved along the same lines as the following theorem
states:

\begin{theorem}
Whatever \ the initial coin density matrix and whatever the
unitary coin reshuffling matrix for the $\varepsilon _{V^{2}}$
quantum random walk, if the asymptotic characteristic function
$h(\phi )$ is not constant then the normalized random variable
$Y^{/}=\frac{Y-\mu }{\sqrt{2}\sigma }$ is distributed according to
the distribution of eq. (\ref{asymp_pdf})$.$
\end{theorem}

This theorem is to be understood as a quantum $\varepsilon _{V^{2}}$ version
of the De Moivre-Laplace theorem for the normal approximation of the
binomial distribution. \ In fact the corresponding theorem one would obtain,
if the quantum system followed the $\varepsilon _{V}$ evolution, would be
precisely the De Moivre-Laplace theorem. In the case of the $\varepsilon
_{V^{n+1}}$ QRW the asymptotic characteristic function is of the form $%
h(\phi )=\mu -\sum_{m=1}^{n}A_{m}\cos 2m(\phi +\alpha _{m}),$
where $A_{m}$ and $\alpha _{m}$ depend on the coin state and the
reshuffling matrix $U.$ If we define $Y_{m}=\cos 2m(\phi +\alpha
_{m})$ then it is not difficult to show that it is distributed
according to eq. (\ref{asymp_pdf}) for any $m.$ Hence we obtain
the following proposition:

\begin{proposition}
If \ the \ asymptotic characteristic function is not constant then the
normalized asymptotic position random variable $Y$ is written as $Y=\mu
+\sum_{m=1}^{n}A_{m}Y_{m},$where each $Y_{m}$ is distributed according to
eq. (\ref{asymp_pdf}), with $\sigma _{Y}^{2}=\frac{1}{2}%
\sum_{m=1}^{n}A_{m}^{2}.$ (It should be noted however that the rv's $Y_{m}$
are strongly correlated). \
\end{proposition}

\section{Cavity Driven Quantum Random Walk}

To probe the effect of initial coin state upon the long time
behaviour of quantum walker system we introduce an operational way
to exercise control on that coin state. If the coin, which is
taken to be a two level atom, is prepared through its interaction
with a single electromagnetic (EM) mode optical cavity on
resonance and if the coin-cavity interaction is assumed to be of
Jaynes-Cummings model (JCM), (see \cite{ellinas2}, and references
therein, also for the case of more general models), then its
dynamic state is described as
\begin{equation}
 \varepsilon
_{U}(\rho _{C})=Tr_{f}U(t)(\rho _{C}\bigotimes \rho
_{f})U(t)^{\dagger }=S_{0}\rho _{C}S_{0}^{\dagger }+S_{1}\rho
_{C}S_{1}^{\dagger }.
\end{equation}%
The CPTP map $\varepsilon _{U}$, describes a stimulate transition
of the coin quantum system due to its interaction with the cavity
mode, and is determined by the unitary evolution $U(t)$ of JCM,
and the initial field state which is taken to be a pure state
$\rho _{f}=|f\rangle \langle f|,$ where $f$ stands for the cavity
photon number. The state of atomic coin system after crossing the
cavity is described by the reduced density matrix given above,
where $t$ stands for the crossing time. The general case of an
initial sharp number state $|f\rangle =|r\rangle ,$ $r=0,1,2,...,$
leads to the reduced coin density matrix
\begin{equation}
\varepsilon _{U}(\rho _{C})=A_{1}\rho _{C}A_{1}^{\dagger }+A_{2}\rho
_{C}A_{2}^{\dagger }+A_{3}\rho _{C}A_{3}^{\dagger }
\end{equation}%
with generators
\begin{eqnarray}
A_{1} &=&\left(
\begin{array}{cc}
\cos (\lambda t\sqrt{r+1}) & 0 \\
0 & \cos (\lambda t\sqrt{r})%
\end{array}%
\right) ,\text{ }A_{2}=\left(
\begin{array}{cc}
0 & 0 \\
\sin (\lambda t\sqrt{r+1}) & 0%
\end{array}%
\right) ,\text{ }  \notag \\
A_{3} &=&\left(
\begin{array}{cc}
0 & \sin (\lambda t\sqrt{r}) \\
0 & 0%
\end{array}%
\right) .
\end{eqnarray}%
The trace preservation of this map requires that $A_{1}^{\dagger
}A_{1}+A_{2}^{\dagger }A_{2}+A_{3}^{\dagger }A_{3}=\mathbf{1}.$

Consider the special case $\rho _{C}=\left\vert c\right\rangle
\left\langle c\right\vert $ where $|c\rangle =\cos \chi |+\rangle
+i\sin \chi |-\rangle ,$ that leads to symmetric walk about the
origin. \ After the coin crosses the JCM cavity, its state becomes
$\varepsilon _{U}(\rho _{C}),$ where
\begin{eqnarray}
\varepsilon _{U}(\rho _{C}) &=&\frac{1}{2}\mathbf{1}+\frac{1}{2}\sin (2\chi
)\cos \left( \ \lambda \sqrt{r+1})t\right) \cos \left( \ \lambda \sqrt{r}%
t\right) \sigma _{2}  \notag \\
&&+\frac{1}{2}\left[ \cos \left( 2\lambda \sqrt{r+1})t\right) \cos ^{2}\chi
-\cos \left( 2\lambda \sqrt{r}t\right) \sin ^{2}\chi \right] \sigma _{3}.
\label{eu}
\end{eqnarray}

Suppose now the prepared atomic coin is used to drive a quantum random walk
according to the $\varepsilon _{V^{2}}$ model for $U=e^{i\frac{\pi }{4}%
\sigma _{2}}$. \ In this \ case the asymptotic characteristic function reads
\begin{eqnarray}
h(\phi ;\chi ,t) &=&\left[ -\cos \left( 2\lambda t\sqrt{r+1})\right) \cos
^{2}\chi +\cos \left( 2\lambda t\sqrt{r}\right) \sin ^{2}\chi \right] \cos
(2\phi )  \notag \\
&&+\left[ \sin (2\chi )\cos \left( \ \lambda t\sqrt{r+1})\right) \cos \left(
\ \lambda t\sqrt{r}\right) \right] \sin (2\phi ).  \label{h_2}
\end{eqnarray}%
To evaluate the limit probability distribution function, we rewrite the last
equation as $h(\phi ;\chi ,t)=C(\chi ,t)\cos [2\phi -\Lambda (\chi ,t)]$,
where we have introduced the functions $A(\chi ,t)=-\cos \left( 2\lambda t%
\sqrt{r+1}\right) \cos ^{2}\chi +\cos \left( 2\lambda t\sqrt{r}\right) \sin
^{2}\chi ,$ $B(\chi ,t)=\sin (2\chi )\cos \left( \ \lambda t\sqrt{r+1}%
\right) \cos \left( \ \lambda t\sqrt{r}\right) $, from which we
define the two functions $C(\chi ,t)=\sqrt{A(\chi ,t)^{2}+B(\chi
,t)^{2}}$\ and $\tan \Lambda (t)=\frac{B(\chi ,t)}{A(\chi ,t)}$. \
If we define the rv $Y=h(\phi ;t),$ then the normalized asymptotic
walker position distribution becomes
\begin{equation}
P(y;\chi ,t)=\frac{1}{\pi \sqrt{C(\chi ,t)^{2}-y^{2}}},\text{ }-1\leq y\leq
1.  \label{pdf_2}
\end{equation}%
This pdf depends on the crossing time $t$ and on the initial coin state
through $\chi $. \ The mean and standard deviation derived from the above
pdf are respectively
\begin{eqnarray}
\mu &=&\lim_{n\rightarrow \infty }\langle \frac{L}{n}\rangle
_{n}=\int\limits_{0}^{2\pi }h(\phi ;\chi ,t)\frac{d\phi }{2\pi }=0, \\
\sigma (\chi ,t)^{2} &=&\lim_{n\rightarrow \infty }\langle \left( \frac{L}{n}%
\right) ^{2}\rangle _{n}\text{ }=\int\limits_{0}^{2\pi }h(\phi ;\chi ,t)^{2}%
\frac{d\phi }{2\pi }=\frac{C(\chi ,t)^{2}}{2}.
\end{eqnarray}%
\

It is useful to observe at this stage\cite{ellinas2}, that the
distribution of eq. (\ref{pdf_2}), is robust under changes in
$\chi ,t$, unless it happens that $C(\chi ,t)=0.$ \
In this case the distribution collapses. \ Inspection of the functions $%
A(\chi ,t),$ $B(\chi ,t)$ and $C(\chi ,t)$ given above reveals that if $\chi
=\{0,\frac{\pi }{2},\pi ,\frac{3\pi }{2}\},$ namely if $|c\rangle
=\{|+\rangle ,i|-\rangle ,-|+\rangle ,-i|-\rangle \}$ respectively, and $t=\{%
\frac{(2k+1)\pi }{4\lambda \sqrt{r+1}},$ $\frac{(2k+1)\pi }{4\lambda \sqrt{r}%
},\frac{(2k+1)\pi }{4\lambda \sqrt{r+1}},$ $\frac{(2k+1)\pi }{4\lambda \sqrt{%
r}}\},$ $k\in Z$, then $C(\chi ,t)=0.$ The specific relations among the four
coin states and interaction times as given above result into only two
different pairs of coin density matrices and interaction times namely, $\rho
_{C}=|+\rangle \langle +|,$ $t=\frac{(2k+1)\pi }{4\lambda \sqrt{r+1}},$ and $%
\rho _{C}=|-\rangle \langle -|,$ $t=\frac{(2k+1)\pi }{4\lambda
\sqrt{r}},$ for which $C(\chi ,t)=0$. If either of these two
conditions occur we say that a \textit{resonance condition }takes
place between the field and the two level atom. \ In this case,
$<L^{2}>_{n}$ $\sim n,$ so we loose the quadratic diffusion time
speed up, characterizing the quantum random walk. In such a case
the asymptotic behaviour of the standard deviation agrees with
that of a classical random walk. More precisely what happens is
that in all
the above cases, the exited coin is in the maximally mixed state $%
\varepsilon _{U}(\rho _{C})=\frac{1}{2}|+\rangle \langle +|+\frac{1}{2}%
|-\rangle \langle -|$ $=\frac{1}{2}\mathbf{1}.$ If this coin system is used
in $V^{2}$ QRW, then the final one-step density matrix for the walker system
becomes
\begin{equation}
\rho _{W}\rightarrow \varepsilon _{V^{2}}(\rho _{W})=\frac{1}{2}\rho _{W}+%
\frac{1}{4}E_{+}^{2}\rho _{W}E_{+}^{\dagger 2}+\frac{1}{4}E_{-}^{2}\rho
_{W}E_{-}^{\dagger 2}.
\end{equation}

Due to the last equation, if $\rho _{W}$ is diagonal initially
then so is finally. Hence we really have a classical one-step
transition that leads to
Gaussian statistics for large $n,$ once we normalize $L$ to $\frac{L}{\sqrt{n%
}}$ . That implies that on resonance the walk becomes fully
classical. This analysis makes obvious the fact that a judicious
choice of the initial coin state permits us to tune the
interaction time in a Jaynes-Cummings cavity, so that we have a
quantum to classical transition on the asymptotic behavior of the
$\varepsilon _{V^{2}}$ quantum random walk. This conclusion makes
the quantum optical experimental investigation of this idea
worthwhile.

\section{Spontaneous Emission in Coin System}

What we are going to study in this section is the effect of a
spontaneously emitting coin system on the QRW evolution. \ We will
assume that the coins that come into contact with the walker
system are corrupted by spontaneous
emission from the state $\left\vert +\right\rangle $ to the state $%
\left\vert -\right\rangle $. \ The state $\left\vert
-\right\rangle $ is assumed to be decay stable, and state
$\left\vert +\right\rangle $ is metastable and has probability
$\gamma $ of decaying to the state $\left\vert -\right\rangle $. \
Let the initial coin state be $\rho _{C}^{0}=p_{0}\left\vert
-\right\rangle \left\langle -\right\vert +p_{1}\left\vert
+\right\rangle \left\langle +\right\vert $, then the effect of
spontaneous emission is to modify this density matrix to $\rho
_{C}^{1}=(p_{0}+\gamma p_{1})\left\vert -\right\rangle
\left\langle -\right\vert +(1-\gamma )p_{1}\left\vert
+\right\rangle \left\langle +\right\vert $. \ This effect can be
captured by the generators\cite{nielsenchuang},
\begin{equation}
\begin{array}{cc}
T_{0}\left\vert -\right\rangle =\left\vert -\right\rangle & T_{1}\left\vert
-\right\rangle =0 \\
T_{0}\left\vert +\right\rangle =\sqrt{1-\gamma }\left\vert +\right\rangle &
T_{1}\left\vert +\right\rangle =\sqrt{\gamma }\left\vert -\right\rangle%
\end{array}%
.
\end{equation}%
In terms of these generators the $\varepsilon_{SE} $ CPTP map that
transforms $\rho
_{C}^{0}$ to $\rho _{C}^{1}$ is%
\begin{equation}
\rho _{C}^{1}=\varepsilon_{SE} (\rho _{C}^{0})=T_{0}\rho
_{C}^{0}T_{0}^{\dagger }+T_{1}\rho _{C}^{0}T_{1}^{\dagger }.
\end{equation}%
Of  course $T_{0}^{\dagger }T_{0}+T_{1}^{\dagger
}T_{1}=\mathbf{1}$. In fact such a spontaneous de-excitation can
happen as result of a stimulated transition as e.g in the case
where the coin passes through an initially empty quantum optical
cavity, (see the case of JCM previously), in which case $\gamma
=\sin ^{2}\lambda t,$  and $ T_{0}=S_{0}(t),$ $T_{1}=S_{1}(t)$,
where $t$ stands for the time spent by the coin inside the empty
cavity. Indeed for the particular initial vacuum state $|f\rangle
=|0\rangle ,$ we obtain the map
\begin{equation}
\rho_{C}^{1}=\varepsilon_{U}(\rho_{C}^{0})= \varepsilon_{SE}
(\rho_{C}^{0})=S_{0}\rho _{C}S_{0}^{\dagger }+S_{1}\rho
_{C}S_{1}^{\dagger }
\end{equation}%
where the so called Kraus generators of that map are
\begin{equation}
S_{0}(t)=\left(
\begin{array}{cc}
\cos (\lambda t) & 0 \\
0 & 1%
\end{array}%
\right) ,\text{ }S_{1}(t)=\left(
\begin{array}{cc}
0 & 0 \\
\sin (\lambda t) & 0%
\end{array}%
\right) ,
\end{equation}%
and satisfy the property $S_{0}^{\dagger }S_{0}+S_{1}^{\dagger }S_{1}=%
\mathbf{1.}$ \ Since in this case no EM field is in the cavity,
$\varepsilon _{U}$ corresponds to spontaneous decay of the excited
atom state (as opposed to stimulated decay if $|f\rangle $ is not
$|0\rangle $), through the emission of a photon.

Suppose now that we have $\varepsilon _{V^{2}}$ evolution with $U=e^{i\frac{%
\pi }{4}\sigma _{2}},$ and that the coin state is not a classical
mixture of states but rather that the coin starts up in the state
$|c\rangle =\cos \chi |+\rangle +i\sin \chi |-\rangle ,$ $\rho
_{C}=\left\vert c\right\rangle \left\langle c\right\vert ,$ then
eq. (\ref{eu}) tells us that
\begin{equation}
\rho_{C}^{1}=\frac{1}{2}\mathbf{1}+\frac{1}{2}\sqrt{1-\gamma }\sin
2\chi \sigma _{2}+\frac{1}{2}[(1-2\gamma )\cos ^{2}\chi -\sin
^{2}\chi ]\sigma _{3}.
\end{equation}%
Since in this case $A(\chi ,\gamma )=-[(1-2\gamma )\cos ^{2}\chi
-\sin ^{2}\chi ]$ and $B(\chi ,\gamma )=\sqrt{1-\gamma }\sin 2\chi
,$ we have that $C(\chi ,\gamma )=\sqrt{[(1-\gamma )\cos 2\chi
-\gamma ]^{2}+(1-\gamma )\sin ^{2}2\chi }$, and the asymptotic
characteristic function is $h(\phi ; \chi ,\gamma )=A(\chi ,\gamma
)\cos 2\phi +B(\chi ,\gamma )\sin 2\phi .$ \ This
gives us that the mean position is zero and that the standard deviation is $%
\sigma (\chi ,\gamma )=\frac{C(\chi ,\gamma )}{\sqrt{2}}.$ \ Note that even
in the case of very strong decay ($\gamma =1$) the asymptotic mean position
of the walker remains zero. \ This is due to the symmetrizing effect of $U$.
\ However the standard deviation of the position depends explicitly on the
decay rate. \ Nevertheless it is not monotonic in the decay rate. \ Since $%
\sigma ^{2}(\chi ,\gamma )=\frac{1}{2}[1-4\gamma (1-\gamma )\cos
^{4}\chi ],$ when $\gamma =0$ or $1$, we have $\sigma
^{2}=\frac{1}{2}$, and this is maximum spread. Furthermore the
spread decreases until $\gamma =\frac{1}{2}$ and then increases
again. \ So we can say that the effect of spontaneous emission is
to \textit{decrease} the spread of the walker position
distribution, however the minimum spread is achieved when the
probability of decay is $\frac{1}{2}.$

\section{Conclusions}

Here we have analysed the asymptotic behaviour of the $\varepsilon
_{V^{k}}$ quantum random walk on Z. \ Firstly, we see that
generically the standard deviation of the position distribution
increases linearly with the number of steps, which is a quadratic
speedup over the classical random walk. \ However there are
resonant quantum random walks in which the asymptotic
characteristic function $h(\phi )$ is constant, which spread
classically. \ Off resonance the normalized position random
variables $\frac{L_{c}^{(n)}}{n}$, are shown to converge weakly to
the random variable $Y=h(\phi ),$ where $ \phi $, is a random
variable whose distribution depends on the initial \textit{walker}
density matrix. \ Furthermore, the asymptotic distribution of the
$\epsilon _{V^{2}}$ model, off resonance is obtained and it is
found that if the position random variable is normalized, then the
asymptotic distribution assumes a universal form given in theorem
4. \ Secondly, it is shown how to drive the asymptotic behaviour
of the $\epsilon _{V^{2}}$ quantum random walk through the use of
a Jaynes-Cummings model interaction that prepares the coin system.
\ It is found that by a judicious choice of the initial coin
state, it is possible to drive the quantum asymptotic behavior of
the walk to classicality by tuning the time spent by coin to cross
the Jaynes-Cummings cavity. \ Finally, the effect of spontaneous
emission of the coin is considered and it is found that in the
particular case of $\varepsilon _{V^{2}}$ evolution with
$U=e^{i\frac{\pi }{4}\sigma _{2}}$, it preserves the symmetry of
the walk, and it actually decreases the spread in the walker
position. \vskip 0.5cm Acknowledgments:  This work was supported
by "Pythagoras II" of EPEAEK research programme.


\end{document}